\documentstyle[amsfonts,graphicx]{europhys}

\def\stars{\bigskip\centerline{***}\medskip}

\newif\ifboo \boofalse

\begin{document}
\euro{}{}{}{}
\Date{}
\shorttitle{M.\ Weigel, W.\ Janke: Universal amplitude-exponent relation for sphere-like lattices}

\title{Universal amplitude-exponent relation for the Ising model on sphere-like
  lattices}

\author{Martin Weigel\footnote{Email: {\tt weigel@itp.uni-leipzig.de}} and Wolfhard
  Janke\footnote{Email: {\tt janke@itp.uni-leipzig.de}}}

\institute{Institut f\"ur Theoretische Physik,
  Universit\"at Leipzig,
  Augustusplatz 10/11, 04109 Leipzig, Germany}

\rec{}{}
\pacs{
\Pacs{75}{10Hk}{Classical spin models}
\Pacs{75}{40Mg}{Numerical simulation studies}
\Pacs{11}{25Hf}{Conformal field theory, algebraic structures}
}

\maketitle

\begin{abstract}
  Conformal field theory predicts finite-size scaling amplitudes of correlation
  lengths universally related to critical exponents on sphere-like, semi-finite
  systems $S^{d-1}\times\mathbb{R}$ of arbitrary dimensionality $d$.  Numerical
  studies have up to now been unable to validate this result due to the intricacies
  of lattice discretisation of such curved spaces. We present a cluster-update Monte
  Carlo study of the Ising model on a three-dimensional geometry using slightly
  irregular lattices that confirms the validity of a linear amplitude-exponent
  relation to high precision.
\end{abstract}

\section{Introduction}
The observation that the symmetry of systems at a critical point goes beyond scale
invariance emerging from the divergence of correlation lengths, and additionally
comprises rotational and translational as well as inversional invariance has greatly
enlarged the scope of exact results that can be derived for such systems on
field-theory grounds.  While customary scaling theory and the real-space
renormalisation ansatz \cite{kadanoff:domb,wegner:domb} give rise to general
relations between the scaling exponents and the concept of universality
\cite{griffiths:70a} of exponents and certain amplitude ratios, conformal field
theory (CFT) \cite{cardy:domb,henkel:book} allows for a classification of models
according to their operator contents, thus providing the {\em values\/} of exponents
and amplitude ratios, and, in special cases, even the amplitudes themselves. By
interpreting the finite size of a system as an additional field in the scaling
ansatz, these considerations apply not only to thermal scaling but as well to the
scaling of observables of finite systems at a critical point in the limit of
diverging system sizes, {\em i.e.\/} to finite-size scaling (FSS) \cite{fisher:74a}.
Concerning universality in FSS, based on renormalisation-group considerations it has
been argued that for semi-finite systems below their upper critical dimension the
correlation lengths (measured in units of the lattice spacing) themselves are
universal quantities \cite{privman:84a}. Exploiting conformal invariance this
statement can be strengthened and the amplitude of this universal correlation-length
scaling can be calculated at least for a special geometry. To see this, consider the
logarithmic map
\begin{equation}
  w=\frac{L}{2 \pi} \ln z, \hspace{0.5cm} z \in \mathbb{C},
  \label{logmap}
\end{equation}
in two dimensions, which wraps the complex plane $\mathbb{C}$ around an
infinite-length cylinder of circumference $L$. Exploiting the facts that the
critical, connected two-point correlation function in the plane is fixed by the
postulate of conformal invariance and that the given map is conformal, the critical
two-point function of any conformally transforming (primary) operator on the cylinder
$S^1\times\mathbb{R}$ follows immediately. In particular, correlations of the
considered operator along the $\mathbb{R}$-axis decline exponentially with a
correlation length of \cite{cardy:84a}
\begin{equation}
  \xi_{\parallel} = \frac{L}{2\pi x},
\label{xi_par}
\end{equation}
where $x$ denotes the scaling dimension of the operator. Thus, one arrives at a FSS
relation including the amplitude $A=1/2\pi x$, implying universality of the
correlation length itself.

Since there is an essential difference between the conformal groups in two
dimensions, where it happens to be infinite dimensional, and higher dimensions, where
it is reduced to a finite-dimensional Lie group, very few CFT results exceeding those
already known from renormalisation-group arguments survive a transition to higher
dimensions.  However, considering the transformation (\ref{logmap}) as a simple
change of coordinates for a moment, rewriting it in polar coordinates allows for a
generalisation to higher dimensions, now mapping ${\mathbb{R}}^d$ to
$S^{d-1}\times\mathbb{R}$. Noting that such a map only acts on the radial part of the
coordinates, leaving the angular part invariant, Cardy \cite{cardy:85a} conjectured
another linear amplitude-exponent relation for the correlation lengths in the
geometry $S^{d-1}\times\mathbb{R}$, namely:
\begin{equation}
  \xi_{\parallel} = \frac{R}{x},
  \label{xi_higher}
\end{equation}
where $R$ denotes the radius of $S^{d-1}$ and $x=x(d)$ is again the scaling
dimension. Note that this includes the $d=2$ result (\ref{xi_par}) with $L=2\pi R$.
This relation, if it can be confirmed, constitutes one of the few exact results for
non-trivial three-dimensional systems.  Since the notion of primarity of a scaling
operator is {\em a priori\/} not well defined for $d>2$, however, it is not entirely
clear to which operators eq.\ (\ref{xi_higher}) should apply. To find the analogues
of primary operators for $d>2$ one might start with the operator product expansion
(OPE), cp.\ the explorative analysis of the operator content of the O($n$) models in
three dimensions that can be found in ref.\ \cite{lang:93a}. Note that this
generalised mapping now connects {\em different\/} geometries, whereas the
two-dimensional mapping can be understood as a meromorphism of the Riemann sphere
onto itself.

Considering alternative approaches, attempts of numerical investigation of the FSS of
systems on such spherical geometries are severely hampered by the fact that there is
only a finite number of triangulations, {\em i.e.\/} regular polyhedra, for each
sphere $S^{d-1}$.  A numerical transfer matrix study for the case of $d=3$ using
Platonic solids as a regular discretisation of the sphere was found to be
inconclusive due to the restriction in system sizes \cite{alcaraz:87a}. Here we
follow a different line and use slightly irregular lattices as sphere discretisations
for $d=3$, exploiting and explicitly probing for the universality of the considered
observables. Considering universality, it might be favourable to study {\em ratios\/}
of correlation lengths which are known to be quite insensitive to lattice
distortions. The attempt to escape such complications by considering the flat
geometry $S^1\times S^1\times\cdots\times\mathbb{R}$ instead surprisingly yields
similar results, which are, however, up to now not connected to CFT calculations
\cite{prl:99a}.

\section{Lattice discretisation}
The most obvious model lattice with spherical topology is given by a cube covered by
a rectangular mesh of lattice points, a lattice type that will be denoted by (C) in
the following. This inevitably introduces irregularities as compared to an idealized
{\em regular\/} (large volume) triangulation of the sphere, which are given by a
concentration of curvature around the cube corners and the defective coordination
number of the corner points (three neighbours instead of four). In view of this
problem several refinements have been proposed such as the replacement of the cube
corners by triangular plaquettes, a construction that hides the coordination number
defects away into the dual lattice, or the projection of the cube geometry (C) on the
sphere $S^2$ by the application of appropriate weight factors to the links between
lattice sites (S) \cite{lang:96a}. This latter lattice can be shown to arrive at the
``right'' continuum field theory in the thermodynamic limit \cite{christ:82a}. On the
other hand, even simpler approximations could be devised such as for example a
``pillow'' resulting from the glueing together of rectangles along their four sides.
Since, at least for bulk quantities, differences between those different choices of
model lattices appear to be small \cite{lang:96a}, especially between (C) and its
projected version (S), we here concentrate on the cubic discretisation (C) which is
computationally much more convenient than (S). As far as universality is concerned,
the {\em ratios\/} of asymptotic correlation lengths are found universal even with
respect to marginal perturbations \cite{yuri:94a}, so that we do not expect to see
artefacts of discretisation defects here. Amplitude universality, however, might be
sensitive to such distortions. Other types of lattice discretisations will be
considered in a forthcoming publication \cite{prep}.

The cube geometry (C) is being realised by six $L\times L$ square lattices
appropriately glued together along their edges.  Since the model lattices including
type (C) are only topologically equivalent to spheres one has to conceive a
definition of effective radii to enable a proper FSS analysis in the sense of eq.\ 
(\ref{xi_higher}). Assigning a unit volume to each lattice site, to each pair of
bonds, or to each individual square of the model lattice, respectively, on counting
these entities one arrives at areas of
\begin{equation}
  A = \left\{
    \begin{array}{c@{}ll}
      6L & (L-2)+8 & {\rm ``sites"}, \\
      6L & (L-2)+6 & {\rm ``bonds", ``squares"},
      \label{radii}
    \end{array}
  \right.
\end{equation}
from which the effective radii result via the relation $R=\sqrt{A/{4\pi}}$. Since for
a ``quadrangulation'' one obviously has
\begin{equation}
4\#({\rm squares})=2\#({\rm bonds}),
\end{equation}
the ``bonds'' and ``squares'' definitions coincide, but differ from the ``sites''
definition by a constant shift, leading to a slightly different approach to the
leading FSS behaviour.

\section{Model and simulation}
We consider a classical, ferromagnetic, nearest-neighbour Ising model with
Hamiltonian
\begin{equation}
  {\cal H}=-J\sum_{\langle i,j\rangle}\sigma_i \sigma_j,\;\;\; \sigma_i=\pm 1.
\end{equation}
The spins reside on lattices compound of the cubical sphere discretisation (C) with
edge lengths $L$ times a further linear lattice direction of length $L_z$. We apply
periodic boundary conditions in $z$-direction to eliminate surface effects in
modelling the infinite $\mathbb{R}$ part. To control the effect of finite $L_z$ still
present we enforce the condition $L_z/\xi_\parallel\gg 1$ in a systematic way, {\em
  i.e.\/} we scale $L_z$ proportionally to the radii $R$ and thus (to leading order)
proportionally to $L$. We use a ratio of $L_z/\xi_\parallel\approx 15$ which
empirically proves sufficient to let the effect of finite $L_z$ drop far below the
threshold of statistical fluctuations. Simulations were done at a quite precise
recent estimate for the Ising model critical coupling in three dimensions
\cite{talapov:96}, $\beta_c=0.221\,654\,4(3)$; the effect of critical coupling
uncertainty was checked by a temperature reweighting technique and found negligible
compared to statistical errors. For the FSS analysis we performed single-cluster
Monte Carlo simulations of nine systems with cube edge lengths ranging from $L=4$ to
$L=12$, corresponding to effective radii of $R\approx2$ to $R\approx8$, and collected
up to about $8\cdot 10^6$ approximately independent measurements after an initial
equilibration phase. Taking the scaled longitudinal sizes $L_z$ into account, system
sizes thus range up to about $3\cdot10^5$ spins.

Within the framework of CFT, the densities of magnetisation and energy turn out to be
primary operators of spin models in two dimensions. Taking this as a hint for higher
dimensions, we thus determine the longitudinal correlation lengths of these two
observables to check whether Cardy's conjecture eq.\ (\ref{xi_higher}) holds in the
discretised case. We start measuring the corresponding connected, longitudinal
correlation functions $G^{c,\parallel}_{\sigma/\epsilon}(z)$, maximally enhancing the
statistics by appropriate averaging and projection techniques, for details see
\cite{conform-long}.  We find that for the extraction of the correlation lengths from
this information at a level of high precision the customary way of fitting the
correlation function data to the functional form
$G(z)^{c,\parallel}=a\,\exp(-z/\xi_\parallel)+b$ for large distances $z$ has severe
drawbacks.  Due to the statistical nature of the data one is not allowed to assume
$b=0$, {\em i.e.\/} the theoretical limit for infinite-length time series, {\em a
  priori\/} for finite-length measurements and is thus forced to use intrinsically
unstable non-linear fits. Therefore, we resort to a difference-ratio type estimator
of the form
\begin{equation}
  \hat{\xi}_\parallel(z)=
  \Delta{\left[\ln\frac{G^{c,\parallel}(z)-G^{c,\parallel}(z-\Delta)}
      {G^{c,\parallel}(z+\Delta)-G^{c,\parallel}(z)}\right]}^{-1},
  \label{diffmethoddelta}
\end{equation}
which eliminates the inconvenient constants $a$ and $b$ above. The parameter
$\Delta\ge 1$ controls the signal-noise ratio of the estimator and is being adapted
for a minimum statistical error. To find final estimates for the correlation lengths
$\xi_\parallel(R)$ of each system, we average the estimates $\hat{\xi_\parallel}(z)$
over a regime of distances $z$ delimited by short distance deviations from the purely
exponential behaviour and exploding fluctuations in the limit of very large
distances. The averaging is done in a statistically optimized way entailing an
expensive analysis of variances and covariances as well as autocorrelations, based on
the ``jackknife'' resampling technique \cite{conform-long}.

\section{Results}
\begin{figure}[t]
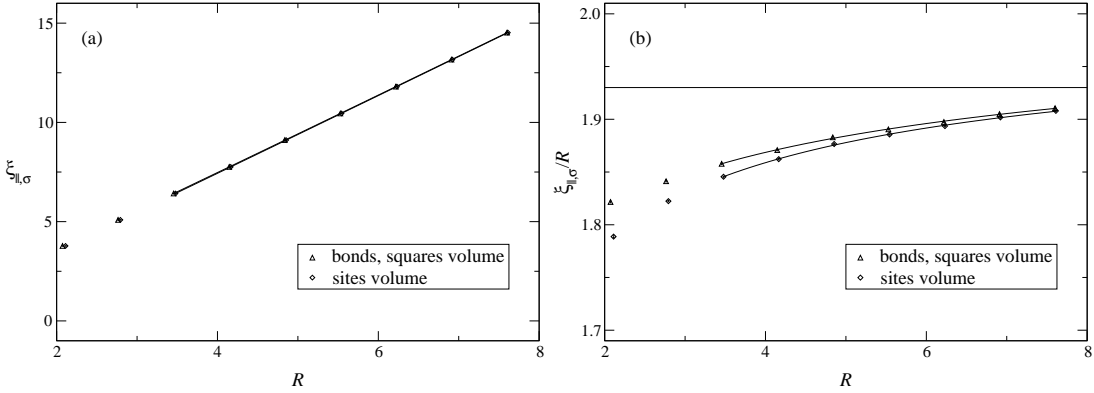

  \mbox{\includegraphics[angle=-90,clip=true,keepaspectratio=true,width=17pc]{scale.eps}
    \includegraphics[angle=-90,clip=true,keepaspectratio=true,width=17pc]{amp.eps}}
  \caption{(a) FSS plot for the spin-spin correlation length $\xi_{\parallel,\sigma}(R)$.
    (b) Scaling of the amplitudes $\xi_{\parallel,\sigma}/R$. The continuous lines
    show fits described in the text and the horizontal line indicates the conjectured
    amplitude $A^{\rm conj}_\sigma=1.9298(13)$.}
  \label{scale_fig}
\end{figure}

Collecting the final estimates for the correlation lengths
$\xi_{\parallel,\sigma/\epsilon}$ of the densities of magnetisation and energy we
arrive at the finite-size results shown in fig.\ \ref{scale_fig}(a), which scale
linearly to leading order in the effective radii $R$ defined above. Differences
between the two radii definitions can hardly be distinguished at this scale. Plotting
the effective amplitudes $\xi_\parallel/R$, however, reveals the presence of
corrections to the leading scaling behaviour that are clearly resolvable at the given
level of accuracy, cp.\ fig.\ \ref{scale_fig}(b). We thus fit our data to a FSS
ansatz including a leading correction term,
\begin{equation}
  \xi_\parallel(R)=AR+BR^{\alpha},
  \label{fitform}
\end{equation}
treating $\alpha$ as an additional fit parameter that thus constitutes an {\em
  effective} correction exponent including parts of the higher-order corrections as
well. This higher-order effect is kept small by successively removing points from the
small $R$ end while monitoring the quality-of-fit parameter $Q$. The range of points
used is indicated by the range of the fit lines in fig.\ \ref{scale_fig}. Combining
the fits to the data for the two different radii definitions we arrive at final
estimates for the leading correlation length scaling amplitudes and their ratio of
\begin{equation}
  \begin{array}{l}
    A_\sigma=1.996(20), \\ A_\epsilon=0.710(38), \\ A_\sigma/A_\epsilon=2.81(15).
  \end{array}
\end{equation}
Using weighted averages of recent estimates for the critical exponents of the
three-dimensional Ising model \cite{conform-long}, namely $\nu=0.63005(18)$ and
$\gamma=1.23717(28)$, the conjectured amplitudes for comparison are
\begin{equation}
  \begin{array}{l}
    A_\sigma^{\rm conj}=1/x_\sigma=2/(d-\gamma/\nu)=1.9298(13), \\
    A_\epsilon^{\rm conj}=1/x_\epsilon=1/(d-1/\nu)=0.70780(23), \\
    A_\sigma^{\rm conj}/A_\epsilon^{\rm conj}=x_\epsilon/x_\sigma=2.7264(13),
  \end{array}
\end{equation}
where $d=3$ is understood, cp.\ eq.\ (\ref{xi_higher}). Comparing the amplitude {\em
  ratios}, which are known to be more robust with respect to lattice inhomogeneities
\cite{yuri:94a,conform-long}, we find very good agreement between the simulation and
Cardy's conjecture. The same holds true for the amplitude of the energy-energy
correlation length whose estimate is naturally less precise than that of the
spin-spin correlation length. For the latter we still obtain reasonable agreement,
but with a deviation of the numerical estimate towards larger values which is on the
edge of statistical significance. To check whether this result indicates a systematic
deviation due to the irregularity of the used lattice discretisation (C), simulations
for the other discretisations mentioned above in the introduction should be performed
\cite{prep}.

\section{Conclusions}
Approximating the sphere $S^2$ by a cube covered by a rectangular mesh of lattice
points, we find that Cardy's conjecture of a linear amplitude-exponent relation in
the FSS of correlation lengths on the sphere-like geometry $S^{d-1}\times\mathbb{R}$
holds true for the case of the three-dimensional Ising model and the densities of
magnetisation and energy. In terms of the analogy to CFT in two dimensions the
considered operators are thus analogues of ``primary'' operators in three dimensions.
The ratio of leading scaling amplitudes does, at the given level of accuracy, not
show any effect of the irregularity of lattice discretisation; for the amplitude of
the spin-spin correlation length further simulations using different discretisation
schemes are necessary to judge the extent of universality that is being obeyed.  Our
results indicate that the geometry $S^2\times\mathbb{R}$ considered here is closer to
the original cylinder $S^1\times\mathbb{R}$ (as far as the prevalence of universal
features of correlation lengths scaling is concerned) than the toroidal geometry
$T^2\times\mathbb{R}$ which is not conformally flat, but nevertheless exhibits a
scaling law quite similar to that found in two dimensions
\cite{prl:99a,conform-long}.

\stars

MW gratefully acknowledges support by the ``Deutsche Forschungsgemeinschaft''
through the Graduiertenkolleg ``Quantenfeldtheorie''.


\vskip-12pt

\begin{thebibliography}{10}

\bibitem{kadanoff:domb}
{\sc Kadanoff L. P.}, in {\em Phase Transitions and Critical Phenomena\/},
  edited by {\sc C.~Domb} and {\sc M. S. Green}, vol.~{\bf 5A} (Academic Press,
  New York) 1976.

\bibitem{wegner:domb}
{\sc Wegner F. J.}, in {\em Phase Transitions and Critical Phenomena\/}, edited
  by {\sc C.~Domb} and {\sc M. S. Green}, vol.~{\bf 6} (Academic Press, New
  York) 1976.

\bibitem{griffiths:70a}
{\sc Griffiths R. B.}, {\em Phys.\ Rev.\ Lett.\/}, {\bf 26} (1970) 1479.

\bibitem{cardy:domb}
{\sc Cardy J. L.}, in {\em Phase Transitions and Critical Phenomena\/}, edited
  by {\sc C.~Domb} and {\sc J. L. Lebowitz}, vol.~{\bf 11} (Academic Press,
  London) 1987, p.~55.

\bibitem{henkel:book}
{\sc Henkel M.}, {\em Conformal Invariance and Critical Phenomena\/} (Springer,
  Berlin/Heidelberg/New York) 1999.

\bibitem{fisher:74a}
{\sc Fisher M. E.}, {\em Rev.\ Mod.\ Phys.\/}, {\bf 46} (1974) 597.

\bibitem{privman:84a}
{\sc Privman V.} and {\sc Fisher M. E.}, {\em Phys.\ Rev.\ B\/}, {\bf 30}
  (1984) 322.

\bibitem{cardy:84a}
{\sc Cardy J. L.}, {\em J.\ Phys.\ A\/}, {\bf 17} (1984) L385.

\bibitem{cardy:85a}
{\sc Cardy J. L.}, {\em J.\ Phys.\ A\/}, {\bf 18} (1985) L757.

\bibitem{lang:93a}
{\sc Lang K.} and {\sc R{\"u}hl W.}, {\em Nucl.\ Phys.\ B\/}, {\bf 402} (1993)
  573.

\bibitem{alcaraz:87a}
{\sc Alcaraz F. C.} and {\sc Herrmann H. J.}, {\em J.\ Phys.\ A\/}, {\bf 20}
  (1987) 5735.

\bibitem{prl:99a}
{\sc Weigel M.} and {\sc Janke W.}, {\em Phys.\ Rev.\ Lett.\/}, {\bf 82} (1999)
  2318.

\bibitem{lang:96a}
{\sc Hoelbling C.} and {\sc Lang C. B.}, {\em Phys.\ Rev.\ B\/}, {\bf 54}
  (1996) 3434.

\bibitem{christ:82a}
{\sc Christ N. H.}, {\sc Friedberg R.} and {\sc Lee T. D.}, {\em Nucl.\ Phys.\
  B\/}, {\bf 210} (1982) 337.

\bibitem{yuri:94a}
{\sc Yurishchev M. A.}, {\em Phys.\ Rev.\ B\/}, {\bf 50} (1994) 13533.

\bibitem{prep}
{\sc Weigel M.} and {\sc Janke W.}, to be published.

\bibitem{talapov:96}
{\sc Talapov A. L.} and {\sc Bl{\"o}te H. W. J.}, {\em J.\ Phys.\ A\/}, {\bf
  29} (1996) 5727.

\bibitem{conform-long}
{\sc Weigel M.} and {\sc Janke W.}, cond-mat/0003124 preprint,
 {\em Phys.\ Rev.\ B}, {\bf 62} (2000), in print.

\end{thebibliography}
\end{document}